\documentclass[doublecol]{epl2} 
\usepackage{subfig}
\usepackage{amsmath}
\graphicspath{{../submission/}}

\begin{document}

\title{The robustness of interdependent clustered networks (25 Sep)}

\author{Xuqing Huang\inst{1} Shuai Shao\inst{1} Huijuan Wang\inst{2,1}
  Sergey V. Buldyrev\inst{1,3} H. Eugene Stanley\inst{1} and Shlomo Havlin\inst{4}}   

\institute{                    
  \inst{1} Center for Polymer Studies and Department of Physics, Boston
  University, Boston, MA 02215 USA\\ 
  \inst{2} Faculty of Electrical Engineering, Mathematics and Computer
  Science,~Delft University of Technology, 2628 CD, Delft, The
  Netherlands\\ 
  \inst{3} Department of Physics,~Yeshiva University, New York, NY 10033 USA\\
  \inst{4} Department of Physics, Bar-Ilan
  University, 52900 Ramat-Gan, Israel 
}
\pacs{89.75.-k}{Complex systems}
\pacs{64.60.ah}{Percolation}
\pacs{64.60.aq}{Networks}

\abstract{It was recently found that cascading failures can cause the
  abrupt breakdown of a system of interdependent networks. Using the
  percolation method developed for single clustered networks by Newman
  [Phys. Rev. Lett. {\bf 103}, 058701 (2009)], we develop an analytical
  method for studying how clustering within the networks of a system of
  interdependent networks affects the system's robustness. We find that
  clustering significantly increases the vulnerability of the system,
  which is represented by the increased value of the percolation
  threshold $p_c$ in interdependent networks.}

\maketitle

\section{Introduction}

In a system of interdependent networks, the functioning of nodes in one
network is dependent upon the functioning of nodes in other networks
of the system. The failure of nodes in one network can cause nodes in
other networks to fail, which in turn can cause further damage to the
first network, leading to cascading failures and catastrophic
consequences. Power blackouts across entire countries have been caused
by cascading failures between the interdependent communication and power
grid systems \cite{Rosato2008,US-Canada-report}. Because infrastructures
in our modern society are becoming increasingly interdependent,
understanding how systemic robustness is affected by these
interdependencies is essential if we are to design infrastructures that
are resilient
\cite{Peerenboom2001,Rinaldi2001,Yagan2011,Vespignani2010}. In addition
to research carried out on specific systems
\cite{Zimmerman2005,Mendonca2006,Robert2008,Reed2009,Bagheri2009,Mansson2009,
Johansson2010,Bashan2012, Li2012, BashanSpatial2012},
a mathematical framework \cite{Sergey2010} and its generalizations
\cite{Parshani2010,Huang2011,Shao2011} have been developed
recently. These studies use a percolation approach to analyze a system
of two or more interdependent networks subject to cascading failure
\cite{GaoNP2011,Gao2011}. It was found that interdependent networks are
significantly more vulnerable than their stand-alone counterparts. The
dynamics of cascading failure are strongly affected by the structure
patterns of network components and by the interaction between
networks. This research has focused almost exclusively on random
interdependent networks in which clustering within component networks is
small or approaches zero. Clustering quantifies the propensity for two
neighbors of the same vertex to also be neighbors of each other, forming
triangle-shaped configurations in the network
\cite{Watts1998,Serrano2006,Newman2003}. Unlike random networks in
which there is very little or no clustering, real-world networks exhibit
significant clustering. Recent studies have shown that, for single networks, 
both bond percolation and site percolation in clustered networks have 
higher epidemic thresholds compared to the unclustered networks
\cite{Newman2009,Miller2009,Gleeson2009,Gleeson2010,Hackett2011,Molina2012}.

Here we present a mathematical framework for understanding how the
robustness of interdependent networks is affected by clustering within
the network components. We extend the percolation method developed by
Newman~\cite{Newman2009} for single clustered networks to coupled
clustered networks. We find that interdependent networks that exhibit
significant clustering are more vulnerable to random node failure than
networks without significant clustering.  We are able to simplify our
interdependent networks model---without losing its general
applicability---by reducing its size to two networks, A and B, each
having the same number of nodes $N$. The $N$ nodes in A and B have
bidirectional dependency links to each other, establishing a one-to-one
correspondence. Thus the functioning of a node in network A depends on
the functioning of the corresponding node in network B and vice versa.
Each network is defined by a joint distribution $P_{st}$ (generating
function $G_0(x,y)=\sum_{s,t=0}^\infty P_{st}x^sy^t$) that specifies the
fraction of nodes connected to $s$ single edges and $t$ triangles
\cite{Newman2009}. The conventional degree of each node is thus
$k=s+2t$. The clustering coefficient $c$ is
\begin{align}
c&=\frac{3\times(\mbox{number of triangles in network})} {\mbox{number
    of connected triples}} \notag
\\ &=\frac{N\sum_{st}tP_{st}}{N\sum_{k}\left(\begin{array}{c}
    k\\2 \end{array} \right)P_k}.
\label{clustering_def}
\end{align}

\section{Site Percolation of single clustered networks}

We begin by studying the generating function of remaining nodes after a
fraction of $(1-p)$ nodes is randomly removed from one clustered network.
After the nodes are removed, we define $t'_i$ to be the number of
triangles of which node $i$ is a part, $d'_i$ to be the number of single
edges that form triangles prior to attack, and $n'_i$ to be the number
of stand-alone single edges prior to attack. This network is thus
defined by the joint distribution $P_{n',t',d'}$.  The probability that
a node has $n'$ single edges from single edges is the sum of all the
probabilities that nodes with more than $n'$ single edges will have
exactly $n'$ edges remaining, which is
$Q_1(n')\equiv\sum\limits_{s=n'}^\infty \left(\begin{array}{c}
  s\\n' \end{array} \right)p^{n'}(1-p)^{s-n'}$. Similarly, the
probability that a node has $t'$ triangles is the sum of all the
probabilities that nodes with more than $t'$ triangles will have exactly
$t'$ triangles remaining. Since the probability that a triangle will
survive is $p^2$, the sum is $Q_2(t')\equiv\sum\limits_{t=t'}^\infty
\left(\begin{array}{c} t\\t' \end{array}
\right)p^{2t'}(1-p^2)^{t-t'}$. The probability that a triangle corner
will have one edge broken is $\frac{2p(1-p)}{1-p^2}$ and the probability
that it will have both edges broken is $\frac{(1-p)^2}{1-p^2}$. Thus the
probability that a node had $d'$ single edges forming triangles prior to
their destruction is $Q_3(d')\equiv\left(\begin{array}{c}
  t-t'\\d' \end{array}
\right)[\frac{2p(1-p)}{1-p^2}]^{d'}[\frac{(1-p)^2}{1-p^2}]^{t-t'-d'}$.
Combining these three, we have the corresponding generating function
\begin{align}
&G(x,y,z,p)=\sum\limits_{n',t',d'}P_{n',t',d'}x^{n'}y^{t'}z^{d'} \notag \\
&=\sum\limits_{n'=0}^\infty x^{n'}Q_1(n') 
\sum\limits_{t'=0}^\infty y^{t'}Q_2(t')\sum\limits_{d'=0}^{t-t'}z^{d'}Q_3(d')P_{s,t}\notag \\
&=G_0(xp+1-p, yp^2+2zp(1-p)+(1-p)^2).
\label{G(x,y,z,p)}
\end{align}

We define $s'=n'+d'$ to be the total number of single links of a node
after attack. The joint degree distribution after attack is $P'_{s',t'}$
which satisfies $P'_{s',t'}=\sum_{n'=0}^{s'} P_{n',t',d'}$, with
$d'=s'-n'$.  The generating function of $P'_{s',t'}$ is
\begin{align}
&G_0(x,y,p)=\sum_{s',t'}P'_{s',t'}x^{s'}y^{t'}\notag \\
&=\sum_{s'=0}^\infty \sum_{n'=0}^{s'}\sum_{t'}P_{n',t',d'} x^{s'}y^{t'}\notag\\
&=\sum_{n',d',t'}P_{n',t',d'}x^{n'} y^{t'} x^{d'} \notag \\
&=G(x,y,x,p).
\end{align}
Therefore, the generating function of the remaining network after attack is 
\begin{align}
G_0(x,y,p)=G_0(xp+1-p, yp^2+2xp(1-p)+(1-p)^2).
\end{align} 
The size of the giant component $g(p)$ of the remaining network
according to Ref.~\cite{Newman2009} is
\begin{align}
g(p)=1-G_0(u,v^2,p),
\end{align}
where
\begin{align}
&u=G_q(u,v^2,p),\\ \notag
&v=G_r(u,v^2,p), \notag
\end{align}
and $G_q(x,y,p)=\frac{1}{\mu }\frac{\partial{G_0(x,y,p)}}{\partial{x}}$,
$G_r(x,y,p)=\frac{1}{\nu }\frac{\partial{G_0(x,y,p)}}{\partial{y}}$ where 
$\mu$ and $\nu $ are the average number of single links and
triangles per node, respectively.

\begin{figure}[h!]
	\includegraphics[width=0.4\textwidth]{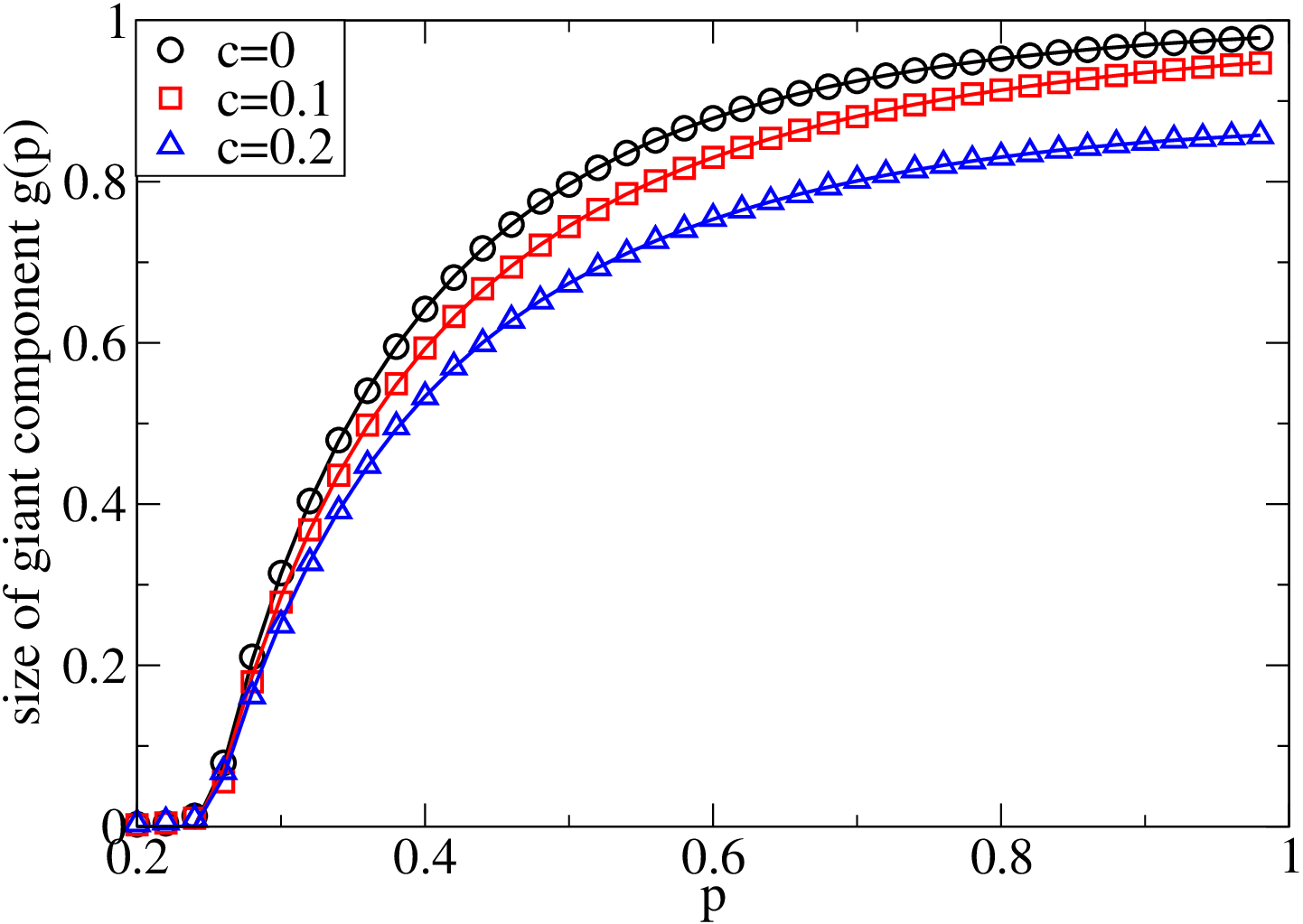}
	\caption{Size of giant component $g(p)$ in single networks with
          degree distribution Eq.~(\ref{double_poisson_pdf}) and average
          degree $\langle k \rangle=4$, as a function of $p$, the
          fraction of remaining nodes after random removal of nodes.
          Curves are from theory Eq.~\ref{giant_component}, symbols are from
          simulation.}
	\label{fig_single_percolation}
\end{figure}

As an example, consider the case when $(1-p)$ fraction of nodes are
removed randomly from a network with doubly Poisson degree distribution
\begin{align}
P_{st}=e^{-\mu}\frac{\mu^s}{s!}e^{-\nu}\frac{\nu^t}{t!},
\label{double_poisson_pdf}
\end{align}
where the parameters $\mu$ and $\nu$ are the average numbers of single
edges and triangles per vertex, respectively.  According to
Eq.~(\ref{clustering_def}), the clustering coefficient is
$c=\frac{2\nu}{2\nu+(\mu+2\nu)^2}$.  Then,
$G_0(x,y)=e^{\mu(x-1)}e^{\nu(y-1)}$ and
$G_0(x,y,p)=G_q(x,y,p)=G_r(x,y,p)=e^{[\mu p+2p(1-p)\nu](x-1)}e^{\nu
  p^2(y-1)}$, and $u=v=1-g(p)$, leading to
\begin{align}
g(p)=1-e^{[\mu p+2p(1-p)\nu]g(p)}e^{\nu p^2(g(p)^2-2g(p))}.
\label{giant_component}
\end{align}
This equation is a closed-form solution for the giant component $g(p)$
and can be solved numerically. The critical case appears when the
derivatives of the both sides of Eq.~(\ref{giant_component}) are
equal. That leads to the critical condition $\langle k \rangle p_c=1$,
which is independent of clustering. 
However the degree distribution of the doubly Poisson model changes 
as we keep the average degree and change the clustering coefficient. 
When the degree distribution is fixed, the critical threshold actually 
increases as clustering increases ~\cite{Gleeson2010, Hackett2011}.
Furthermore, Fig.~\ref{fig_single_percolation} shows the resulting giant
component as a function of $p$. Note that single networks with
higher clustering have smaller giant components.

\section{Degree-Degree Correlation}

When constructing clustering in a network, it is usually impossible to
avoid generating degree-degree correlations. To better understand the
effect of clustering on degree-degree correlations, we present an
analytical expression of degree correlation as a function of the
clustering coefficient for a doubly Poisson-clustered network---see
Eq.~(\ref{double_poisson_pdf}).

The degree-degree correlation \cite{MieghemEPJB2010} can be expressed as
\begin{equation}
\rho _{D}=\frac{N_{1}N_{3}-N_{2}^{2}}{N_{1}\sum \limits_{i=1}^{N}d_{i}^{3} N_{2}^{2}}
\label{assortativitydef}
\end{equation}
where $N_{m}$ is the total number of $m$ hop walks between all possible
node pairs $(i,j)$ including cases $i=j$.

The generating function of the degree of a node in the network is
$\sum\limits_{s,t=0}^{\infty}P_{st}z^{s+2t}=G_{0}(z,z^{2})$. Let
$q_{st}$ be the fraction of nodes with $s$ single edges and $t$
triangles that are reached by traversing a random single link, where $s$
includes the traversed link and $r_{st}$ is the fraction of nodes with
$s$ single edges and $t$ triangles reached by traversing a link of a
triangle, $q_{st}=\frac{sP_{s,t}}{\langle
  s\rangle},r_{st}=\frac{tP_{s,t}}{\langle t\rangle }$. Their
corresponding generating functions are $G_{q}(x,y)=\frac{1}{\langle
  s\rangle}\frac{\partial G_{0}(x,y)}{\partial x}x$ and
$G_{r}(x,y)=\frac{1}{\langle t\rangle }\frac{\partial
  G_{0}(x,y)}{\partial y}y$. Moreover,
$N_{3}=\sum\limits_{i}\sum\limits_{j}a_{ij}N_{2}(j)$, where $N_{2}(j)$
is the total number of two-hop walks starting from node $j$. The number
of three-hop walks from a node $i$ is equal to the total number of
two-hop walks starting from all of its neighbors. Thus,
$N_{3}=\sum\limits_{j}k_{j}N_{2}(j)$, where the number of two-hop walks
starting from a node $j$ with degree $k_{j}$ will be counted $k_{j}$
times in $N_{3}$.  Equivalently,
$N_{3}=N\sum\limits_{st}(s+2t)P_{s,t}N_{2}(s,t)$, where $N_{2}(s,t)$ is
the number of two hop walks from a node with $s$ single edges and $t$
triangles. The generating function of the number of single edges and of
triangles reached in two hops from a random node is
$G_{2}(x,y)=\sum\limits_{st}P_{s,t}\cdot G_{q}^{s}(x,y)\cdot
G_{r}^{2t}(x,y)$.  The generating function of the total number of links
and of triangles reached within three hops starting from all nodes is
$G_{3}(x,y)=N\sum\limits_{st}P_{s,t}\cdot \left( G_{q}(x,y)\right)
^{s(s+2t)}\cdot \left(G_{r}(x,y)\right) ^{2t(s+2t)}$.  The number
$N_{k}$ of $k$-hop walks can be approximated by its mean in a large
network
\begin{eqnarray*}
N_{1} &=&N\langle k\rangle, \\
N_{2} &=&N\frac{\partial G_{2}}{\partial x}\left\vert
x=1,y=1\right.+2N\frac{\partial G_{2}}{\partial y}\left\vert
x=1,y=1\right. \\ 
N_{3} &=&\frac{\partial G_{3}}{\partial x}\left\vert
x=1,y=1\right. +2\frac{\partial G_{3}}{\partial y}\left\vert
x=1,y=1\right.  
\end{eqnarray*}

When both $s$ and $t$ follow a Poisson distribution,
\begin{eqnarray*}
G_{0}(x,y) &=&e^{\mu (x-1)}e^{\nu (y-1)} \\
G_{q}(x,y) &=&G_{0}(x,y)x \\
G_{r}(x,y) &=&G_{0}(x,y)y.
\end{eqnarray*}
In this case,
\begin{eqnarray*}
N_{1} &=&N\langle k\rangle  \\
N_{2} &=&N\left\langle k\right\rangle \left( \frac{\left\langle
k\right\rangle }{1-c}+1\right)  \\
N_{3} &=&\left( \langle k\rangle ^{3}+2\langle k\rangle ^{2}+4\nu \langle
k\rangle +\langle k\rangle +6\nu \right) N\text{ } \\
\sum\limits_{i=1}^{N}d_{i}^{3} &=&\left( \left\langle k\right\rangle
^{3}+3\left\langle k\right\rangle ^{2}+(6\nu +1)\left\langle
k\right\rangle +6\nu \right) N,
\end{eqnarray*}
which together with Eq.~(\ref{assortativitydef}) leads to
\begin{equation}
\rho_D=\frac{c-c^2-\langle k\rangle c^2}{1-c+\langle k\rangle c-2\langle
  k\rangle c^2}, 
\label{correlation}
\end{equation}
where c is the clustering coefficient, Eq.~(\ref{clustering_def}).

Figure~\ref{fig_clustering_correlation} shows the relation between the
degree correlation and the clustering coefficient $c$ for a Poissonian
network [see Eq.~(\ref{double_poisson_pdf})], for two given average
degrees ($\langle k \rangle=3$ and 4). The figure shows a positive
degree-degree correlation across the entire range, which means the model
is assortative \cite{Gleeson2010}.  The degree-degree correlation
increases until $c$ achieves half of its maximum and then decreases to
zero when $c$ reaches its maximum. When $c$ is 0 or the maximum, the
nodes connect to either all single links or all triangles, respectively.

\section{Percolation on Interdependent Clustered Networks}

To study how clustering within interdependent networks affects a
system's robustness, we apply the interdependent networks framework
\cite{Sergey2010}. In interdependent networks A and B, a fraction $(1-p)$ of
nodes is first removed from network A. Then the size of the giant
components of networks A and B in each cascading failure step is defined
to be $p_1$, $p_2$, ..., $p_n$, which are calculated iteratively
\begin{equation}
\begin{array}{l}
p_n=\mu_{n-1}g_A(\mu_{n-1}), \mbox{n is odd,}\\
p_n=\mu_{n}g_B(\mu_n), \mbox{n is even,}
\end{array}
\label{stages}
\end{equation}
where $\mu_0=p$ and $\mu_n$ are intermediate variables that satisfy
\begin{equation}
\begin{array}{l}
\mu_n=pg_A(\mu_{n-1}), \mbox{n is odd,}\\
\mu_n=pg_B(\mu_{n-1}), \mbox{n is even.}
\end{array}
\label{stages2}
\end{equation}
As interdependent networks A and B form a stable mutually-connected giant
component, $n\rightarrow \infty$ and $\mu_n=\mu_{n-2}$, the fraction of
nodes left in the giant component is $p_\infty$. This system satisfies
\begin{equation}
\begin{array}{l}
x=pg_A(y),\\
y=pg_B(x),
\end{array}
\label{Eq11}
\end{equation}
where the two unknown variables $x$ and $y$ can be used to calculate
$p_\infty=xg_B(x)=yg_A(y)$. Eliminating $y$ from these equations, we
obtain a single equation
\begin{equation}
  x=pg_A[pg_B(x)].
  \label{Eq12}
\end{equation}
The critical case ($p=p_c$) emerges when both sides of this equation
have equal derivatives,
\begin{equation}
  1=p^2\frac{dg_A}{dx}[pg_B(x)]\frac{dg_B}{dx}(x)|_{x=x_c,p=p_c},
  \label{Eq13}
\end{equation}
which, together with Eq.~(\ref{Eq12}), yields the solution for $p_c$ and
the critical size of the giant mutually-connected component,
$p_\infty(p_c)=x_cg_B(x_c)$.

Consider for example the case in which each network has doubly-Poisson
degree distributions as in Eq.~(\ref{double_poisson_pdf}). From
Eq.~(\ref{Eq11}), we have $x=p(1-u_A)$, $y=p(1-u_B)$, where
\begin{equation}
\begin{array}{l}
u_A=v_A=e^{[\mu_A y+2y(1-y)\mu_A](u_A-1)+\nu_A p^2(v_A^2-1)},\\ \notag
u_B=v_B=e^{[\mu_B x+2x(1-x)\mu_B](u_B-1)+\nu_B p^2(v_B^2-1)}.\\ \notag
\end{array}
\end{equation}
If the two networks have the same clustering, $\mu\equiv\mu_A=\mu_B$ and
$\nu\equiv\nu_A=\nu_B$, $p_\infty$ is then
\begin{equation}
p_\infty=p(1-e^{\nu p_\infty^2-(\mu+2\nu)p_\infty})^2.
\label{Eq15}
\end{equation}

\begin{figure}[h!]
\subfloat[]{
	\includegraphics[width=0.4\textwidth]{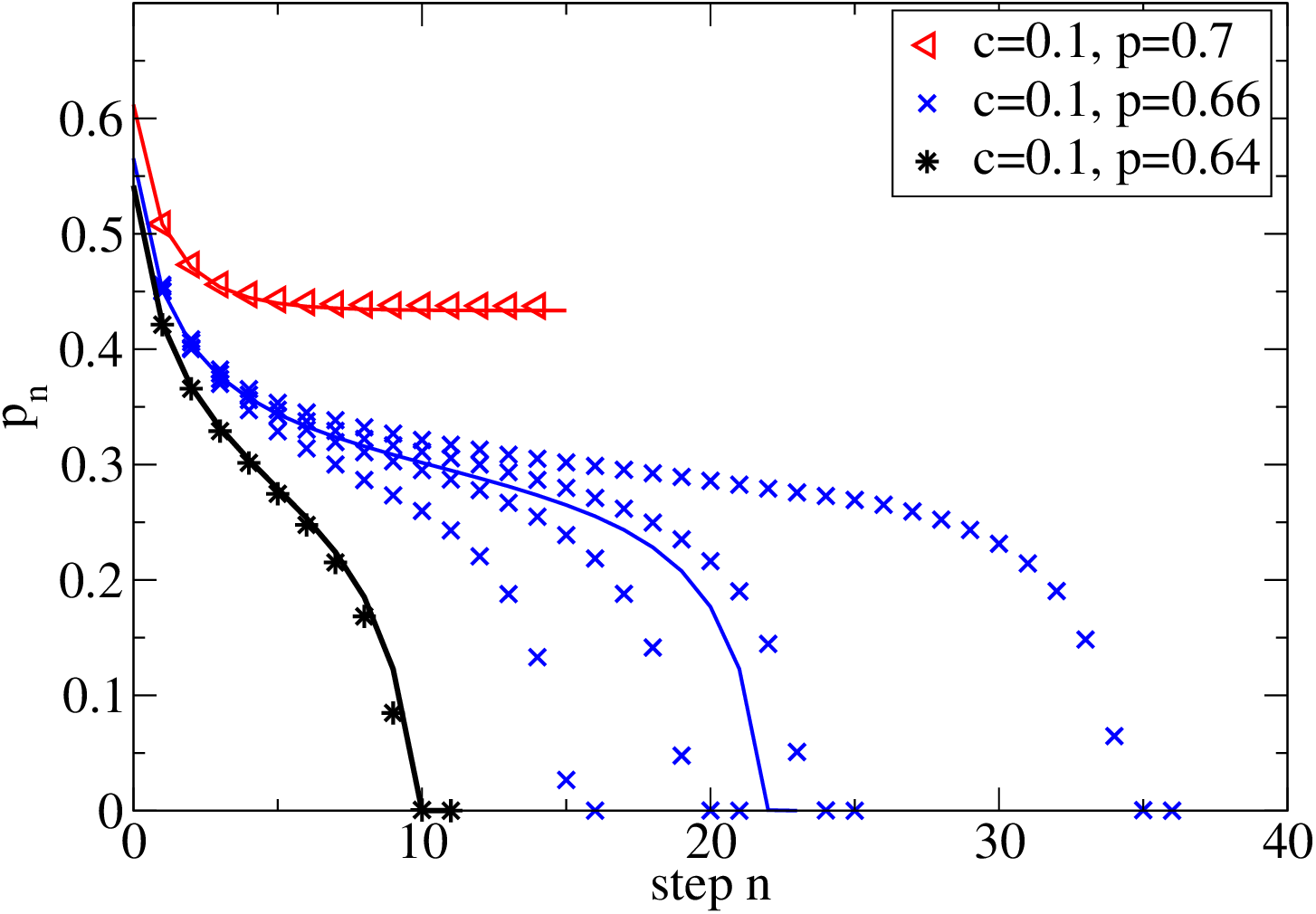}
	\label{steps}}\\
\subfloat[]{
	\includegraphics[width=0.4\textwidth]{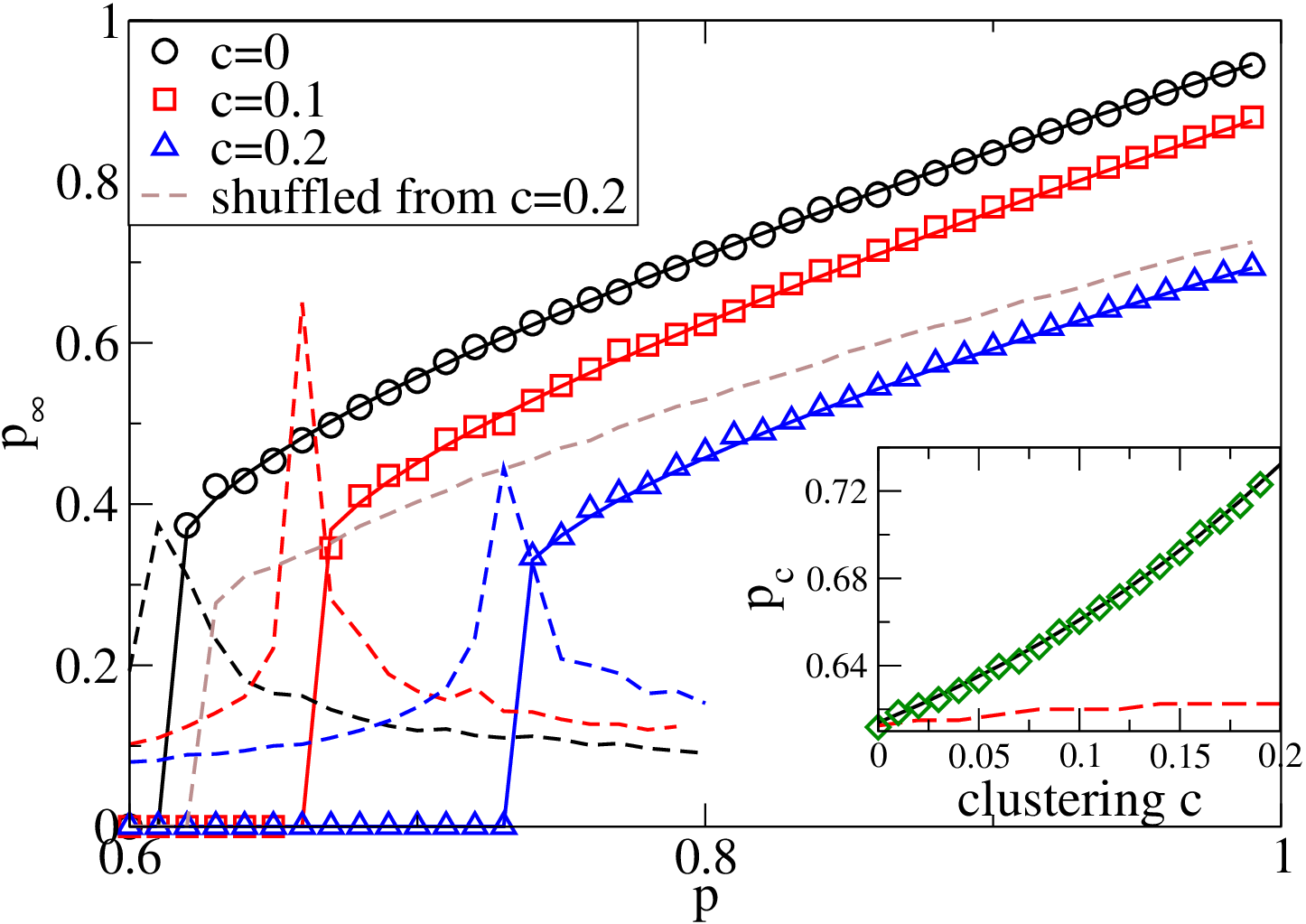}
	\label{fig_inter_percolation_fix_k}}
\caption{
  (a) Size of mutually connected giant component as a function of
  cascading failure steps $n$. Results are for $c=1$, $p=0.64$ (below
  $p_c$), $p=0.66$ (at $p_c$) and $p=0.7$ (above $p_c$). Lines represent
  theory (Eqs.~(\ref{stages}) and (\ref{stages2})) and dots are from
  simulations. Note that at $p_c$ there are large fluctuations.  
  (b) Size of giant component, $p_\infty$, in interdependent networks with
  both networks having clustering via degree distribution
  Eq.~(\ref{double_poisson_pdf}) and average degree $\langle k
  \rangle=4$, as a function of $p$. Dashed lines are number of
  interactions~(NOI) before cascading failure stops
  obtained by simulation \cite{ParshaniPNAS}. The star curve is for shuffled $c=0.2$ network,
  which keeps the same degree distribution but without clustering and without
  degree-degree correlation.
  Inset: Green squares and solid line represents critical thresholds, $p_c$, of
  interdependent networks as a function of clustering coefficient $c$. 
  Red dashed line represents critical threshold of shuffled interdependent 
  networks which originally has clustering coefficient $c$. 
  The shuffled networks have zero clustering and degree-degree
  correlation, but has the same degree distribution as the original clustered networks.
  In all figures, symbols and dashed lines represent simulation, 
  solid curves represent theoretical results.
  }
\label{fig_percolation}
\end{figure}

\begin{figure}[h!]
\subfloat[]{
	\includegraphics[width=0.4\textwidth]{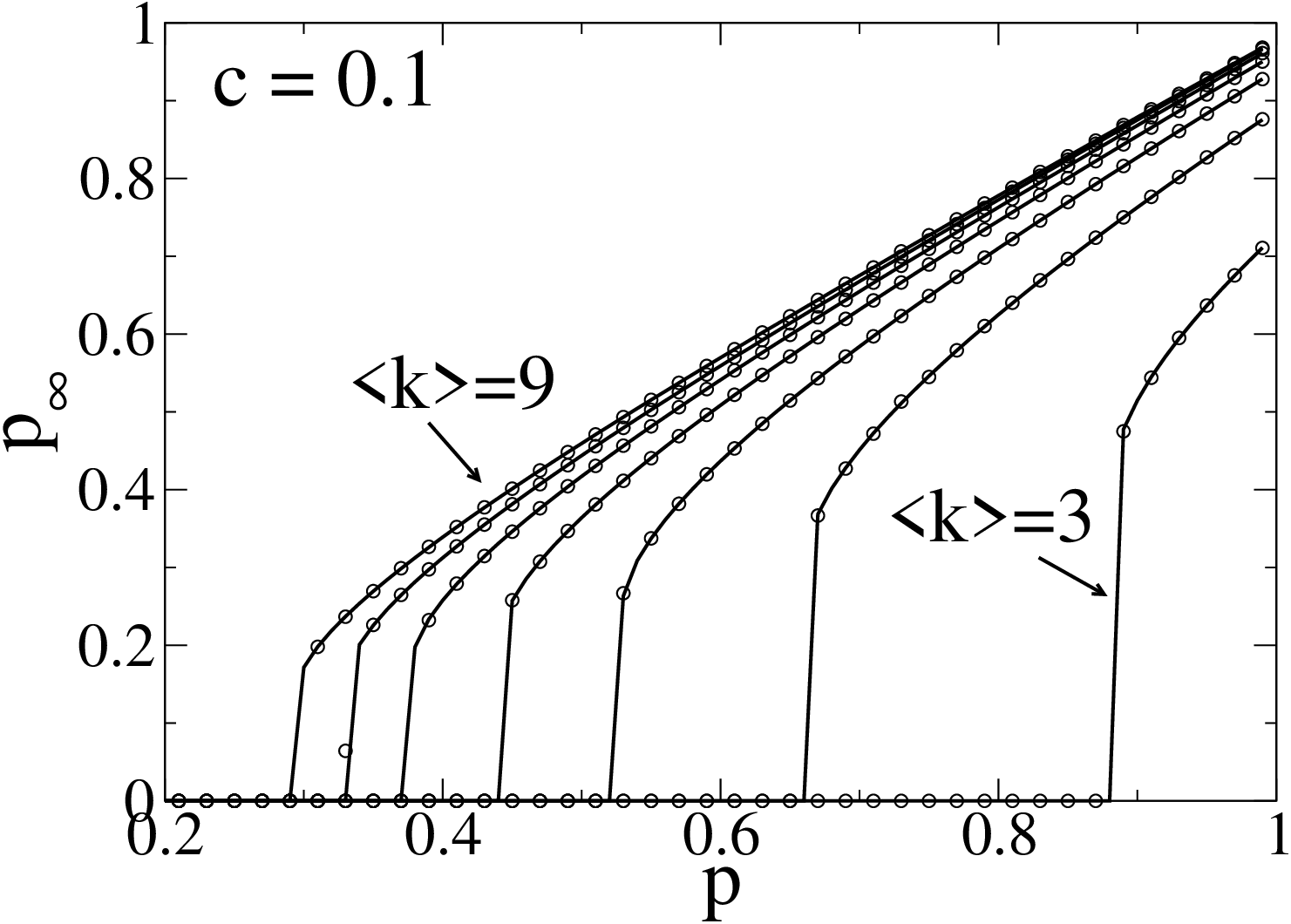}
	\label{fig_inter_percolation_fix_c}}\\
\subfloat[]{
	\includegraphics[width=0.4\textwidth]{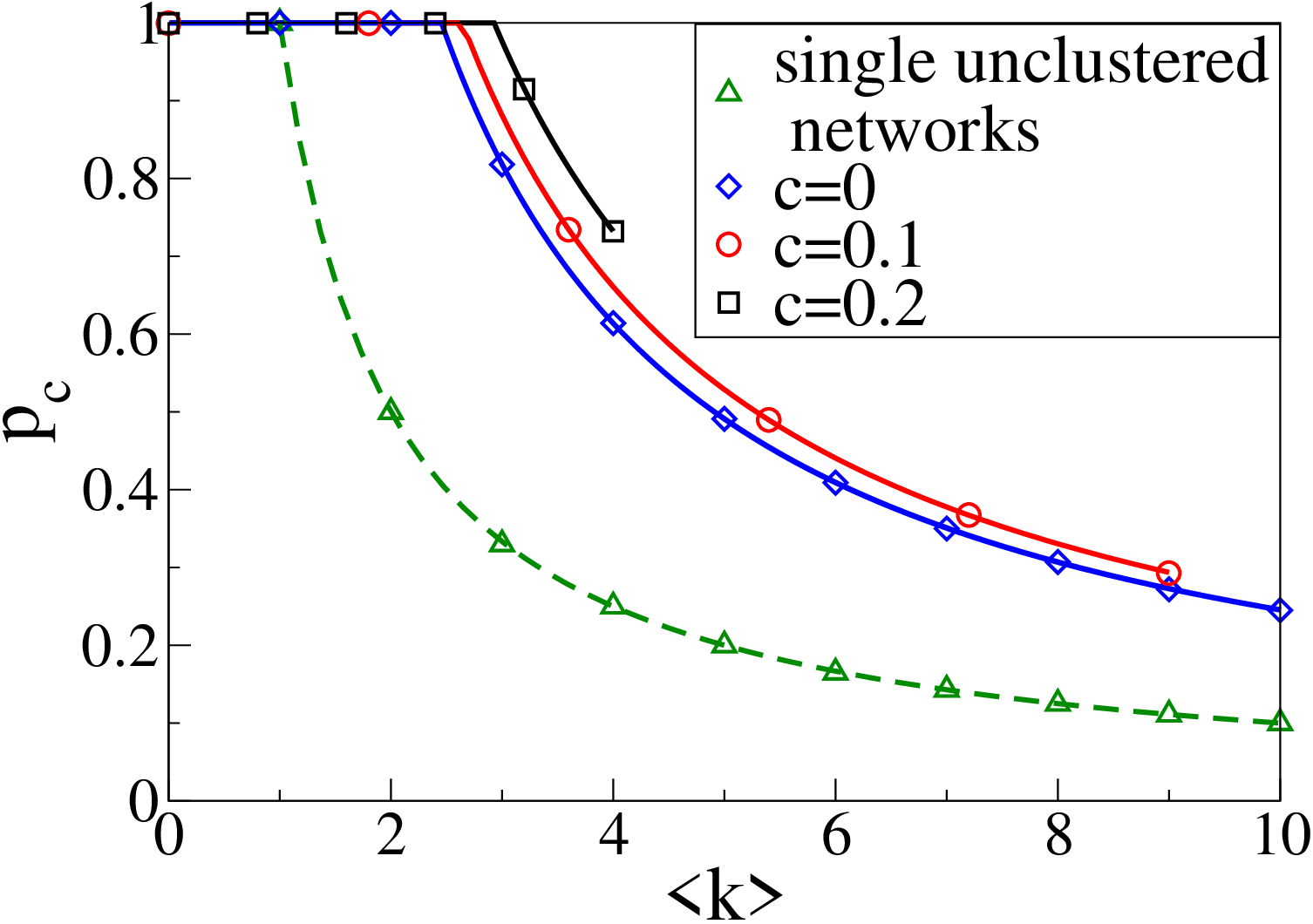}
	\label{pc_c}}
\caption{(a) Size of giant component as a function of $p$ for fixed
  clustering coefficient $c=0.1$ and different average degrees. From
  right to left $\langle k\rangle=3, 4, 5, ..., 9$. (b) Critical
  threshold $p_c$ as a function of average degree for different
  clustering coefficients. The solid curves are for interdependent
  networks and the dashed curve is for single networks. Symbols and curves
  represent simulation and theoretical predictions respectively.}
\label{fig_percolation}
\end{figure}

The giant component, $p_\infty$, for interdependent clustered networks
can thus be obtained by solving Eq.~(\ref{Eq15}). Note that when $\nu=0$
we obtain from Eq.~(\ref{Eq15}) the result obtained in
Ref.~\cite{Sergey2010} for random interdependent ER networks.
Figure~\ref{steps}, using numerical simulation, compares the size of the
giant component after $n$ stages of cascading failure with the
theoretical prediction of Eq.~(\ref{stages}). When $p=0.7$ and $p=0.64$,
which are not near the critical threshold ($p_c=0.6609$), the agreement
with simulation is perfect. Below and near the critical threshold, the
simulation initially agrees with the theoretical prediction but then
deviates for large $n$ due to the random fluctuations of structure in
different realizations~\cite{Sergey2010}. By solving Eq.~(\ref{Eq15}),
we have $p_\infty$ as a function of $p$ in
Fig.~\ref{fig_inter_percolation_fix_k} for a given average degree and
several values of clustering coefficients and in
Fig.~\ref{fig_inter_percolation_fix_c} for a given clustering and for
different average degree values. As the figure shows, when higher
clustering within a network is introduced, the percolation transition
yields a higher value of $p_c$ (see inset of
Fig.~\ref{fig_inter_percolation_fix_k}).

\begin{figure}[h!]
	\includegraphics[width=0.4\textwidth]{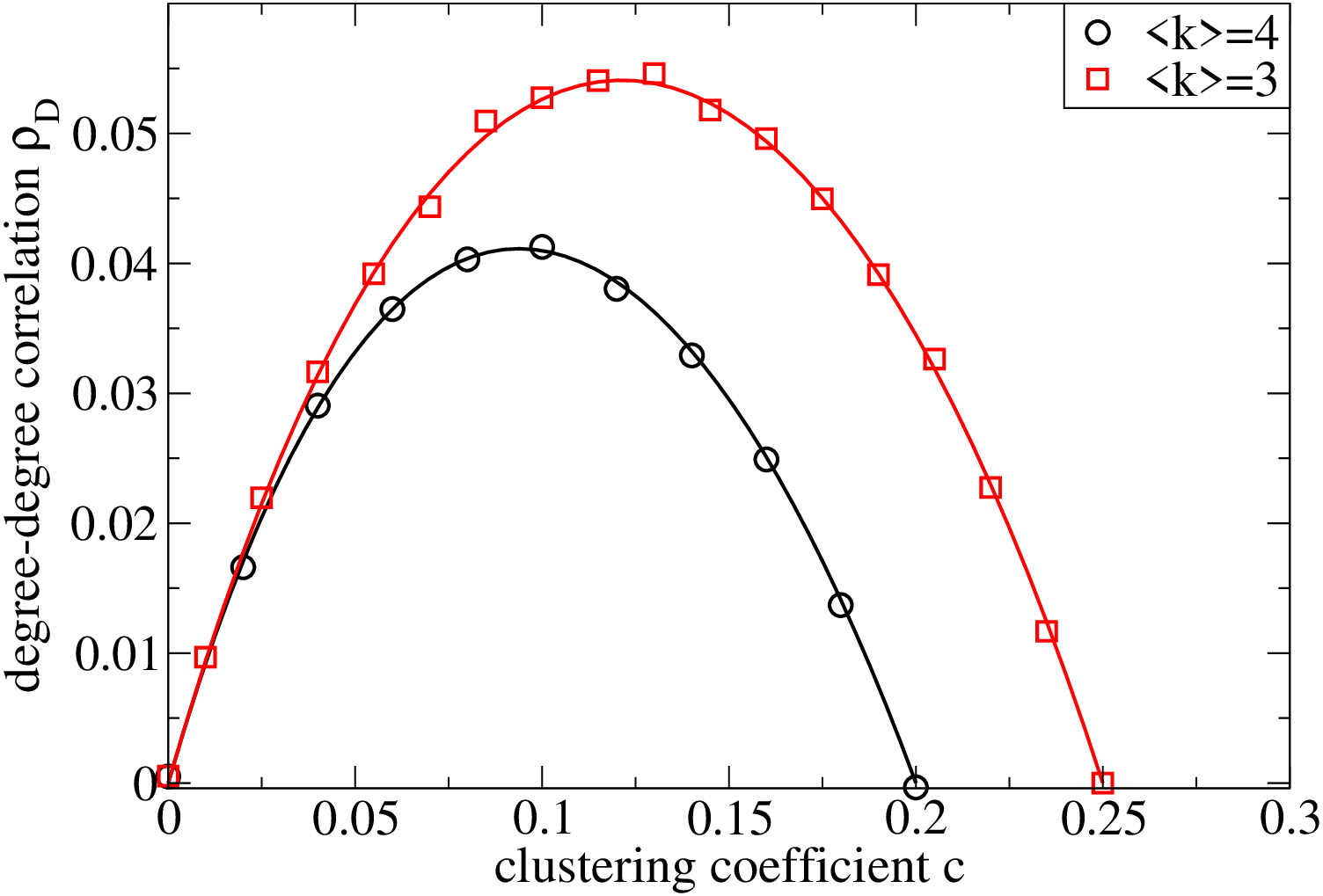}
	\caption{Degree-degrees correlation as a function of the
          clustering coefficient for Poisson network
          (Eq.~(\ref{double_poisson_pdf})) with average degree $\langle
          k\rangle=3$ and 4. Curves are from theory
          (Eq.~\ref{correlation}) and symbols from simulations.}
	\label{fig_clustering_correlation}
\end{figure}

When clustering changes in this doubly Poisson distribution model,
degree distribution and degree-degree correlation also change.  First,
to address the influence of the degree distribution, we study the
critical thresholds of shuffled clustered networks. Shuffled clustered
networks have neither clustering nor degree-degree distribution but keep
the same degree distribution as the original clustered networks.  The
brown dashed curve in Fig.~\ref{fig_inter_percolation_fix_k} represents
the giant component of interdependent shuffled clustered networks with
original clustering $c=0.2$.  The figure shows that the difference in
$p_c$ between the $c=0$ network and the shuffled $c=0.2$ network is only
$0.01$, while the difference between the $c=0$ and the $c=0.2$ networks
is $0.12$.  In addtion, $c=0.2$ clustered networks has no degree-degree
correlation (Fig.~\ref{fig_clustering_correlation}), which means the
$0.12$ shift of $p_c$ is due to clustering and not to a change in degree
distribution.  We also show the critical thresholds of interdependent
shuffled clustered networks as the red dashed line in the inset of
Fig.~\ref{fig_inter_percolation_fix_k}.  Note that the change of degree
distribution barely shifts the critical threshold.  We next discuss the
effect of the degree-degree correlation on the change of critical
threshold.  From Ref.~\cite{Zhou2012}, the degree assortativity alone
monotonously [??? JDM] increases the percolation critical threshold of
interdependent networks.  Because in our case degree-degree correlation
first increases and then decreases (see
Fig.~\ref{fig_clustering_correlation}), while critical the threshold of
interdependent networks increases monotonously [???] as clustering
increases, we conclude that clustering alone increases the value of
$p_c$.  Thus clustering within networks reduces the robustness of
interdependent networks.  This probably occurs because clustered
networks contain some links in triangles that do not contribute to the
giant component, and in each stage of cascading failure the giant
component will be smaller than in the unclustered case.

We also study the effect of the mean degree $\langle k \rangle$ on the
percolation critical point. Figures~\ref{fig_inter_percolation_fix_c}
and \ref{pc_c} both show that, when clustering is fixed, the percolation
critical point of interdependent networks decreases as the average
degree $\langle k\rangle$ of network increases, making the system more
robust.  Figure~\ref{pc_c} also shows that a larger minimum average
degree is needed to maintain the network against collapse without any
node removal as clustering increases.
	
\section{Conclusion and Summary}	

To conclude, based on Newman's single network clustering model, we
present a generating-function formalism solution for site percolation on
both single and interdependent clustered networks. We also derive an
analytical expression, Eq.~(\ref{correlation}), for degree-degree
correlation as a function of the clustering coefficient for a
doubly-Poisson network. Our results help us better understand the effect
of clustering on the percolation of interdependent networks. We discuss
the influence of a change of degree distribution and the degree-degree
correlation associated with clustering in the model on the critical
threshold of interdependent networks and conclude that $p_c$ for
interdependent networks increases when networks are more highly
clustered.

\acknowledgments

We wish to thank ONR Grant\# N00014-09-1-0380, DTRA Grant\#
HDTRA-1-10-1-0014, the European EPIWORK, MULTIPLEX, CONGAS and LINC
projects, DFG, the Next Generation Infrastructures (Bsik) and the Israel
Science Foundation for financial support.

\end{document}